\documentclass[10pt,final,journal,twocolumn]{IEEEtran}
\usepackage{amsmath}
\usepackage{graphicx, epstopdf}
\usepackage{setspace}
\usepackage{acronym}
\usepackage{commath}
\usepackage{mathrsfs}
\usepackage{mathtools}
\usepackage{ifthen}
\usepackage{textcomp}
\usepackage{float}
\usepackage[tight,footnotesize]{subfigure}
\usepackage{cite}
\usepackage{bm}
\usepackage{url}
\usepackage[utf8]{inputenc}
\usepackage{mathtools}
\usepackage{amssymb}
\usepackage{amsthm}
\usepackage{mathrsfs}
\usepackage{bm}
\usepackage{multirow}
\usepackage{multicol}
\usepackage{xspace}
\usepackage{tabularx}

\begin{document}

\title{Securing Autonomous Air Traffic Management: Blockchain Networks Driven by Explainable AI
\thanks{$^{1}$University of Oxford, UK; $^{2}$Cranfield University, UK; $^{3}$Alan Turing Institute, UK; $^{4}$University of Glasgow, UK; $^{5}$TEKTowr, UK; 
$^*$Corresponding Author: weisi.guo@cranfield.ac.uk
The authors wish to acknowledge the UKRI Future Flight Phase 2 Grant: Fly2Plan – Enabling a new model aviation data system-of-systems (70883)}
}
\author{
\IEEEauthorblockN{Louise Axon$^{1}$, Dimitrios Panagiotakopoulos$^{2}$, Samuel Ayo$^{2}$, Carolina Sanchez-Hernandez$^{2}$, Yan Zong$^{2}$,\\
Simon Brown$^{5}$, Lei Zhang$^{4}$, Michael Goldsmith$^{1}$, Sadie Creese$^{1}$, Weisi Guo$^{2,3*}$, }}

\maketitle

\begin{abstract}
Air Traffic Management data systems today are inefficient and not scalable to enable future unmanned systems. Current data is fragmented, siloed, and not easily accessible. There is data conflict, misuse, and eroding levels of trust in provenance and accuracy. With increased autonomy in aviation, Artificially Intelligent (AI) enabled unmanned traffic management (UTM) will be more reliant on secure data from diverse stakeholders. There is an urgent need to develop a secure network that has trustworthy data chains and works with the requirements generated by UTM. Here, we review existing research in 3 key interconnected areas: (1) blockchain development for secure data transfer between competing aviation stakeholders, (2) self-learning networking architectures that distribute consensus to achieve secure air traffic control, (3) explainable AI to build trust with human stakeholders and backpropagate requirements for blockchain and network optimisation. When connected together, this new digital ecosystem blueprint is tailored for safety critical UTM sectors. We motivate the readers with a case study, where a federated learning UTM uses real air traffic and weather data is secured and explained to human operators. This emerging area still requires significant research and development by the community to ensure it can enable future autonomous air mobility. 
\end{abstract}

\begin{IEEEkeywords}
aeronautical telecommunications; blockchain; explainable machine learning; air traffic management; unmanned traffic management;
\end{IEEEkeywords}

\IEEEpeerreviewmaketitle

\section{Introduction}

\begin{figure}[t]
     \centering
     \includegraphics[width=1\linewidth]{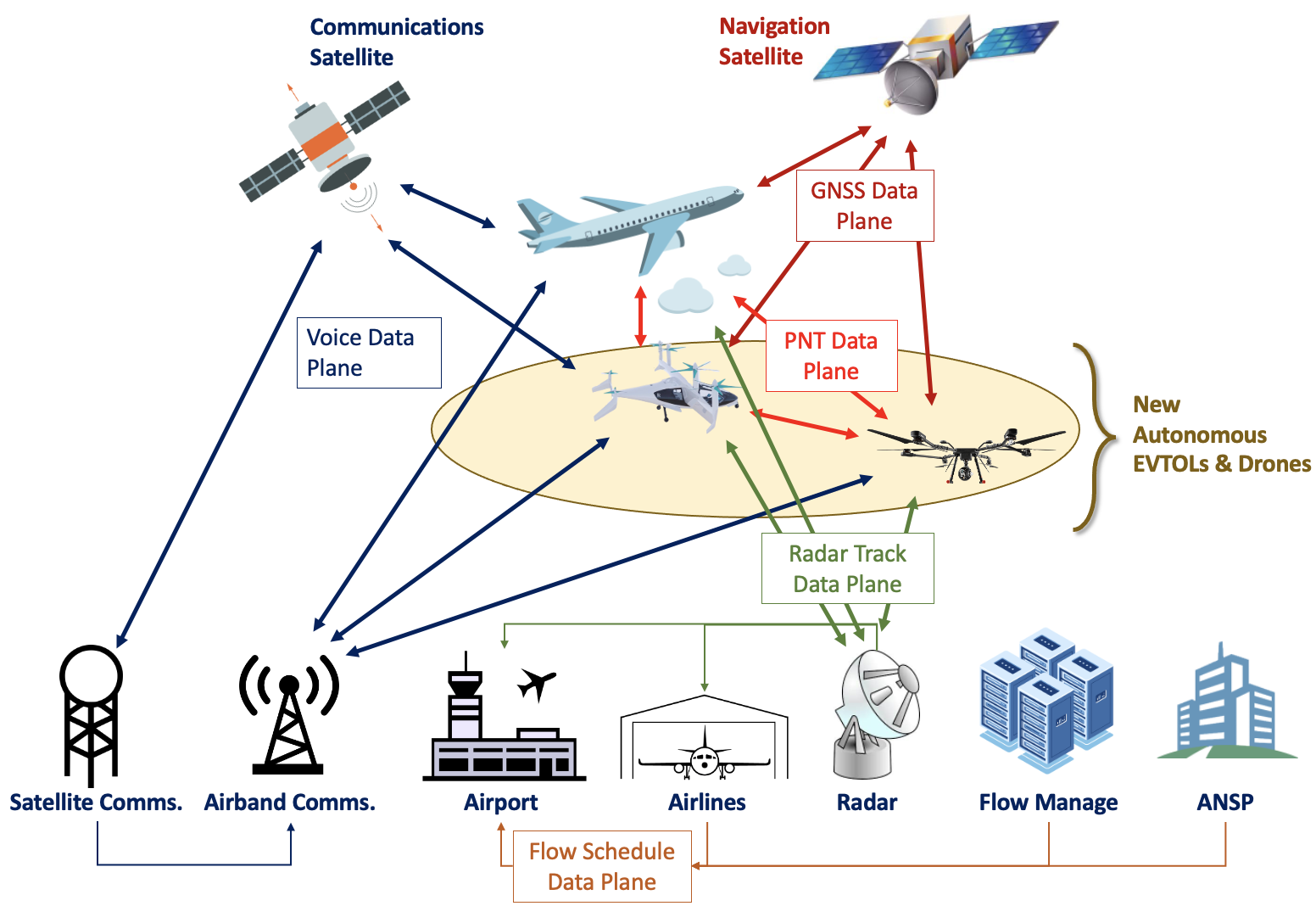}
     \caption{Data planes in air traffic management (ATM): Air Navigation Service Providers (ANSPs) work with airlines and airports to deliver air traffic flow and capacity management. It relies on synchronized and secure PNT data from GNSS, radar, and flight. Large volumes of autonomous drones threaten to disrupt this ecosystem, requiring data sharing across new commercial and private entities and overwhelming traditional traffic flow management operators.}
     \label{fig0}
\end{figure}

The aviation industry shares flight data through internationally-defined standard format messages - see International Civil Aviation Organisation (ICAO). As shown in Fig.\ref{fig0}, Air Navigation Service Providers work with airlines and airports to deliver air traffic flow and capacity management. It relies on synchronized and secure Positioning, Navigation, and Timing (PNT) data from Global Navigation Satellite System (GNSS), radar track data, and flight voice data. The current ecosystem is designed and operated based upon the assumption that aircraft would largely fly airport-to-airport along predetermined routes. This process is no longer fit for purpose, particularly as we enable UTM traffic and transition towards On-Demand Air Mobility.

\subsection{Challenges in Air Traffic Management Networks}

Aviation data is constantly at risk of being eroded and misused, and large volumes of autonomous and spontaneous drone flights threaten to disrupt this data ecosystem:
\begin{itemize}
    \item Data misuse: airlines strip out commercially sensitive fuel consumption detail, resulting in air navigation and airport service providers using low-quality information, frequently causing uncertainty at landing runways. Recent studies indicate that as much as 60\% of flight delays are due to data errors or network in-availability. EASA and Eurocontrol estimates that 1 hour of delay to 1 flight costs \$8-18k. Whilst the bulk of these often happen in bursts (e.g., Alaska ATM virus, British Airways personal data breach, Russian air space satellite data manipulation). Furthermore drone data is often at risk of being misused in a wide range of adversarial and criminal ways, putting aviation safety at risk.
    \item Emerging autonomy: Unmanned aerial systems (UAS) dramatically increase the need for unplanned traffic management at a scale human air traffic controllers cannot cope with. Unmanned aviation is set to unlock £45bn of benefits to the UK economy by 2030, with over €10-15bn of benefits per year estimated for the EU. UAS or drones fly often lower than conventional aircraft, increasing their risk. Without a pilot, they are heavily dependent on reliable and timely data acquisition and sharing to ensure their flight is conducted safely, often using new data sources not provided by the legacy air traffic network.
\end{itemize}
\textbf{Cybersecurity Challenge Example:} Currently, airlines strip out commercially sensitive fuel consumption details, resulting in air traffic management using low-quality information and causing uncertainty at landing queue management. This has a significant safety impact and a lasting economic impact at the airport (60\% of delays are due to data errors or in-availability). EASA and Eurocontrol estimates that 1 hour of delay per flight costs \$8-18k. Whilst the bulk of these cybersecurity issues occur in bursts (e.g., Alaska ATM virus, BA data breach), UTM and UASs’ high dependency on data-driven autonomy puts us at a level of public safety risk unseen before. These challenges highlight the impeding data crisis as we move towards a more autonomous aviation.

\subsection{Current Research Overview \& Gaps}
Whilst the Single European Sky ATM Research (SESAR) programme and American NextGen encourage greater data sharing through System-Wide Information Management (SWIM) \cite{AGData}, this still requires organisations to individually combine and rationalise shared data. Not only is this cumbersome, current research initiatives also fail to accommodate new airspace users operating from non-airports and flying bespoke routes. Emerging unmanned aerial systems (UAS) often use non-aviation positioning or communication channels such as 5G and terrestrial timing distribution services. All of these new and decentralised capabilities must be accommodated in a cyber-secure way, providing high levels of trust and low latency accuracy. 

Modern Distributed-Ledger Techniques (DLT) are distributed over wireless networks offer a potential solution to accommodate the sheer variety of data and participating systems in a coherent trust-ensuring manner \cite{SpaceAirBC, BlockchainAir}. The key systems include the need for explainable AI to enable unmanned traffic management (UTM) \cite{EASA, Collision} to manage increased data complexity and flow scheduling load. This is being actively addressed by emerging EASA frameworks and Eurocontrol standards \cite{EASA} and international R\&D programs (NASA AOSP, FAA NextGen, EU SESAR, Eurocontrol ECHO, UKRI Future Flight). 

Existing review and survey papers have largely focused on either end-user applications of DLT-enabled UASs \cite{9184022} or integration with space and terrestrial 5/6G ecosystems \cite{9631953, 9316453}. What has not been examined is the enabling of the ATM and air traffic control systems which is much more time-sensitive and safety critical. The sheer volume of data between many stakeholders (airlines, airports, ATM, radar,...etc.) makes this challenging. This review paper comes at the back of a major UK Future Flight project (Fly2Plan) and will review the state-of-the-art research in emerging AI Integrated DLT technology to support all airspace management.

\subsection{Novelty \& Contribution: Explainable UTM with Secure Blockchain Data Streams}

\begin{figure}[t]
     \centering
     \includegraphics[width=1\linewidth]{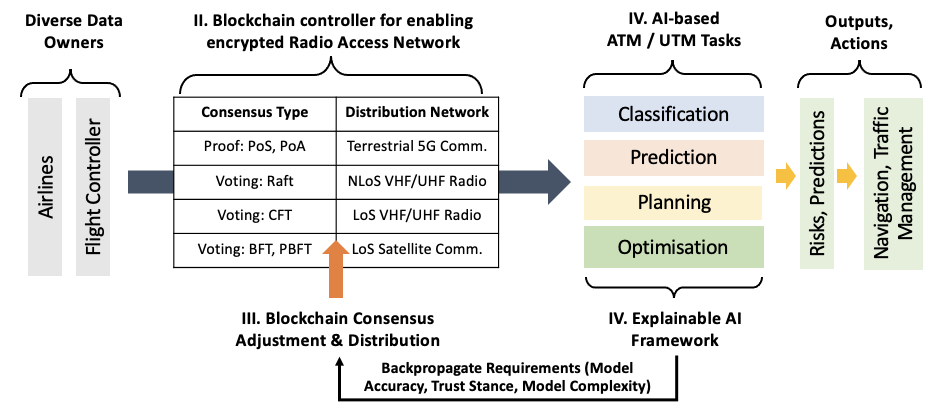}
     \caption{Future air traffic management: AI-based air traffic management driven by secure data chains. Deep learning algorithms use explainable AI (XAI) interfaces to discover the data security requirements. These are backpropagated to the blockchain controllers to distribute consensus via wireless network.}
     \label{fig1}
\end{figure}

UTM will adopt AI algorithms to relief the cognitive burden of human operators at the different entities in Fig.\ref{fig0}. As AI is sensitive to data inputs in different ways that humans are, this places even greater emphasis on dynamic levels of trust and security on the data that informs AI decisions. This can only be achieved if the AI algorithms backpropagate security requirements back to the networked blockchain modules that distribute consensus across the communication network \cite{Lei22}. The review sets out recent advances which foster the development of wireless network architectures that enable the novel collaboration between mission-critical explainable AI and blockchain secured data streams. The review paper is organised as follows (see Fig.\ref{fig1}) with novel contributions in reviewing state-of-the-art research in the following areas:
\begin{enumerate}
    \item Section II: A review of current wireless network architectures used to optimise and disseminate consensus whilst being ATM/UTM context aware.
    \item Section III: A review of blockchain advances for UTM, especially in the context of adaptively setting the aviation security requirements and working between competing commercial stakeholders.
    \item Section IV: A review of explainable AI advances for automated air traffic control, and the backpropagation of trust and security requirements to blockchain and consensus distribution. A federated learning case study with real air traffic and weather data is used to demonstrate securing and explaining UTM. 
    \item Section V: Summary of the overall vision of future automated ATM data networks, as well as open challenges.
\end{enumerate}
On top of the review, we also include our own demonstration case studies using real aviation data. We hope this review enables readers to think about the joint challenges between secure networked AI for mass air autonomy and wireless network design. 

\section{Wireless Consensus Distribution}
\label{section2}

We aim to improve the security and data-driven AI trust in future ATM/UTM communications to address increased unmanned drone usage and alleviate human traffic controller's cognitive burden. This can be achieved between multiple stakeholders (see Fig.\ref{fig0}) using recent advances in distributing consensus in a blockchain (see Fig.\ref{fig1}) \cite{Lei22}. The consensus between all nodes is a critical component of a modern blockchain (detailed in Section III), while the underlying communication network that makes quorums possible is essential. In the case of ATM/UTM, the traffic management sensitive data/decisions (see Section IV) need to be agreed by all traffic participants and regulators in the area through achieving a consensus. Sequentially, all nodes can be updated with latest logs of decisions with non-repudiable records in the blockchain. 

Unlike wired systems, wireless systems bring extra channel uncertainty and scarcity of spectrum, thus entailing different distributed consensus protocols and performance. A complete protocol with privacy preservation that is suitable to UTM needs to be established. Information-theoretic models are needed to bridge the achievable bounds and real performance between wireless channels and DLT. Operationally, malicious attacks through open wireless environment may compromise voting-based consensus, raising the need to consider new threat models to DLT and to develop novel spectrum optimisation for adaptive secure networking. As shown in Fig.\ref{fig0} ATM/UTM scenarios, most traffic participants rely on wireless communications links, e.g., VHF (118-137 MHz), UHF (225-400 MHz), ADS-B (978/1090 MHz), and satellite communications (27/35 GHz), to build up situational awareness and make decisions with additional feedback from primary, secondary radar systems, and other GNSS capabilities. Therefore, the blockchain network, upon the consensus group between all manned/unmanned aircraft, ground control centers and regulators, needs to have the quorum over wireless network.

\begin{table*}[ht]
\centering
\caption{ATM/UTM Areas of Operations: Network, blockchain, and (X)AI Requirements.}
\begin{tabular}{|>{\centering}p{65pt}|>{\centering}p{75pt}>{\centering}p{75pt}>{\centering}p{75pt}>{\centering}p{75pt}>{\centering}p{75pt}|}
		\hline
        ATM/UTM Area & Criticality and (Latency Requirement) & Wireless Channel (Typical Range) & Consensus Type \cite{Lei22} (Section II-III) & UTM AI Algorithms \cite{EASA} (Section IV) & Suitable XAI Methods \cite{LiTrust20} (Section IV) \cr \hline 
       
        Collision Avoidance \cite{Collision} & Safety Critical (100ms) & LoS/NLoS ($>$5km) & Byzantine Fault Tolerance (BFT) & Kalman filter, RNN & Feature: SHAP \cr \hline 
        
        Remote Piloting & Safety Critical ($<$1ms) & NLoS ($<$25km) & Byzantine Fault Tolerance (BFT) & Kalman filter, RNN & Feature: SHAP \cr \hline 
        
        Aircraft Detection \cite{EASA} & Safety Critical ($<$2s) & LoS ($<$200km) & Proof-based consensus (PoX) & Filter banks, CNN \& RNN & Feature: SHAP, DeepLIFT \cr \hline  
        
        Delay Prediction \cite{Dimitri21} & Mission Critical ($<$1min) & LoS ($<$5km) & Proof-based consensus (PoX) & Expert Models, ARMA, RF, GP, RNN & Feature: SHAP \cr \hline  
        
        Navigation Planning & Mission Critical (s) & LoS ($<$350km) & Crash Fault Tolerance (CFT) & Q-Learning, DRL & Feature: DeepLIFT \cr \hline 
        
        Autonomous Landing \cite{EASA} & Safety Critical ($\approx$10ms) & LoS ($<$15km) & Byzantine Fault Tolerance (BFT) & CNN \& Q-Learning / DRL & Feature: DeepLIFT; Summative: TDA \cr \hline 
        
        Air Space Congestion Management \cite{BlockchainAir} & Mission Critical ($\approx$100ms) & LoS ($<$50km) & Voting: Raft or PBFT & Clustering \& Convex Opt. / Feed-forward NN / Bayesian Tree & Summative: LIME, Meijer-G, TDA \cr \hline
\end{tabular}
\label{comparison}
\end{table*}

\subsection{ATM Driven Consensus Type Selection}
The consensus type of ATM blockchain can be decided based on the ATM tasks (full summary in Table I):
\begin{itemize}
    \item \textbf{Airspace Co-Existence:} shared airspace availability information can be updated into the network using Proof-based consensus (PoX), e.g., Proof-of-Stakes (PoS), or Proof-of-Authorities (PoA), as the local aviation information may not have impacts with other regional air traffic participants. 
    \item \textbf{Congestion Management:} a multi-drone congestion advisory message (e.g., near logistics pick up centre) is better handled by the voting-based consensus, e.g., Raft or Practical Byzantine Fault Tolerance (PBFT), which other participants may have their input on whether agreeing the message during the consensus. PBFT is particularly suitable due to robustness against sensing errors in air space usage and aircraft positioning, as well as when under malicious data injection.
    \item \textbf{Collision Avoidance:} in latency sensitive tasks \cite{Collision}, voting-based consensus is preferred, as it can greatly increase the reliability and availability of the ATM. There are two major groups of voting-based consensuses. (1) the Crash Fault Tolerance (CFT) consensus and (2) Byzantine Fault Tolerance (BFT). The CFT consensus assumes all nodes are benign in the network. However, they may crash or be out of service due to the limited reliability and availability of the node. In contrast, BFT handles the network with malicious nodes, producing false or inaccurate network feedback. The CFT can help boost the reliability of ATM decision-making, and BFT can help the ATM have better tolerance to sensor errors and perform error corrections to traffic participants. 
\end{itemize}

\subsection{Safety Critical Consensus Types and Decentralised Optimisation}

Upon the deployment of the consensus, wireless communications is important for the consensus network, as both CFT and BFT require multi-phase messaging to other nodes in the network:
\begin{enumerate}
    \item Crash Fault Tolerance (CFT): the wireless network can be simplified as a MAC layer network, where each node only broadcasts and listens to the open space without a central network controller. Each node may broadcast its message based on its status within the local consensus group. 
    \item Byzantine Fault Tolerance (BFT): it requires a significant number of communications between all nodes when the network grows, which is a challenge for the wireless channel capacity whilst maintaining a minimum latency standard. Hence the MAC layer-only network might be a bottleneck for the consensus performance in large drone swarm situations, as the transmission options are limited to the MAC layer with limited flexibility of broadcast and uni-cast. 
\end{enumerate}

To optimise consensus parameters on a distributed network, where each aviation platform may not have full knowledge of the consensus distribution network, one can use machine learning \cite{Lei22}. The research in \cite{SpaceAirBC} reviewed novel air-ground-space blockchain-assisted virtulalised network architectures. Within that framework, credit-based consensus algorithm in a lightweight vehicular blockchain can be used to securely and immutably trace malicious data access and record data transactions for aviation systems with improved efficiency and security in reaching consensus. The main benefit is that given local aerial platforms have little explicit knowledge of the whole network, reinforcement learning (Q-learning on the voting-based consensus outcomes) is used at each node to optimally schedule the pricing and quality of data sharing strategies for both data contributor and data consumer via trial and error. In contrast, the research in \cite{NeuralBlockChain} present a centralised neural network optimised blockchain-based mobile drone edge computing (MEC), designed to ensure ultra-reliability and provide a flat architecture. The master blockchain is created such that only the pre-existing block information from the respective blockchain of an individual is operated over the neural model to generate the optimized values. 

\subsection{Open Research Areas}

The current active research areas in this space are:
\begin{itemize}
    \item \textbf{Secure information-theoretic bounds:} Establish secure link budget models by investigating how DLT system metrics (e.g., transaction delay, throughput, security threshold) and wireless communication (e.g., delay, throughput, and reliability) can be integrated. Research in the UTM environment can focus on voting based consensuses and cascade compromise scenarios. Recognising that the consensuses’ procedures can be significantly different and communication relevant, the model for each consensus should be built respectively according to current and future aviation networks, e.g., crash based consensus (RAFT) may be used for latency-sensitive tasks; while Byzantine fault tolerant protocols can be used for some high secure cases without stringent latency requirements. 
    \item \textbf{Optimal secure communication resource allocation:} The available communication resource in terms of power and spectrum for the consensus air peer-to-peer network can significantly affect consensus thus DLT performance. For instance, increasing communication transmission power can benefit DLT security through a broader coverage of the broadcasting range since it means faster/safer procedure for consensus. Thus, there is a lack of research on how to minimise the required communication resources subject to required DLT performance. 
    \item \textbf{Zero-knowledge proof (ZKP) based DLT:} The highlight of ZKP is to realise the proof without revealing anything other than the statement is true (i.e., no leakage of the statement-related information). As such, the coupling of ZKP and DLT could be a promising technical route to bridge the gap of sharing sensitive information between airline companies, ATCT, aviation administrations and other stakeholders. An open area of research is to use ZKP to protect the privacy of airline companies based on DLT, which is served as a platform to underpin ZKP protocols using smart contracts to share sensitive information (e.g., remaining fuel for landing optimisation) and ensure the shared information is validated and synchronised among stakeholders.
\end{itemize}

\section{Adversarial Attacks in Blockchain Enabled UTM}
\label{section3}

Blockchain-based systems have the potential to enable data exchange between ATM and competing commercial stakeholders (through consensus) - creating a reliable data ledger (a shared source of “truth”). They also have the potential to meet important requirements on the security of this data. If designed securely, blockchain-based systems can produce data ledgers with high levels of integrity and availability through their core mechanisms. These properties are vital for safety and mission-critical UTM contexts; in particular when there is a risk of lack of data integrity. 

\subsection{Tamper Resistance}
Firstly, a high level of tamper-resistance can be achieved. It is well-understood that the storage of hash-chained data along with resource-limited block proposal makes manipulating stored data extremely difficult. This can help ensure data maintains integrity. For example, a BFT Blockchain can also withstand some malicious or faulty nodes, such that the system can continue to operate correctly, maintaining availability of the data exchange under such circumstances. Furthermore, the availability of data does not rely on the availability of a single or small set of storage machines since the ledger is distributed across multiple nodes who store a copy. However, as is the case with all systems, the security of blockchain-based systems depends on their design and the operational security practices that surround them. The integrity, availability and confidentiality of data in these data chains could be put at risk by various attacks and faults. Detailing all possible attacks is outside the scope of this paper, but some illustrative examples follow.

\subsection{Adversarial Attack Examples}

\subsubsection{Resource Control}
Attacks against consensus algorithms are a risk if an adversary gains control of a sufficiently high level of the network’s \textit{resources}, e.g., mining power in a system that uses Proof-of-Work (PoW) as the block-proposal mechanism. This may enable an adversary to manipulate data (by reverting to a new main chain) or manipulate finalisation (i.e., decide which blocks are finalised into the chain). Gaining this resource may be achieved through a range of approaches, e.g., an infrastructure-level attack (e.g., stealing credentials) to gain control of an endpoint, Sybil attack, Eclipse attack, collusion between nodes, or exploiting a vulnerability in the resource-limitation algorithm (e.g., a hashing vulnerability in the PoW algorithm). The level of resource required is dependent on the block-finalisation algorithm used. One example is achieving the 51\% attack against the probabilistic longest-chain algorithm \cite{ReviewBC}. 

\subsubsection{Coding Vulnerabilities}
Another example is the exploitation of coding vulnerabilities in smart contracts, which could potentially enable adversaries to disable or lock the smart contract, delete or manipulate data, gain unauthorised access to data, post data without authorisation, or restrict legitimate control. Such vulnerabilities could occur accidentally or be inserted deliberately during initial development or updates to smart contracts. Attacks on access-control and identity-management mechanisms in permissioned Blockchain systems, which could enable participation by unintended parties, also need to be considered. Addressing these risks to security requires secure design and coding practices, security testing of systems and effective operational security practice to identify and mitigate the impact of malign behaviour by insiders and outsiders to the network. 

\subsection{Improving Security \& Privacy for UTM}
There has been research into how to integrate additional security and privacy techniques with Blockchain systems in order to achieve necessary security properties. This includes, for example, the use of Trusted Execution Environments (TEEs) for executing smart contracts securely. In the ATM context, while data integrity and availability are clearly critical for safety and efficiency of operations, the confidentiality of certain data or participants is also important in certain cases – for example while this data is sensitive for commercial or security reasons. Alongside restricting access to a set of authorised parties through private, permissioned Blockchains, privacy-preserving techniques can be used to enable necessary operations: Zero-Knowledge proofs (e.g., ZK-SNARK), ring signatures (e.g., Monero), and secure multi-party computation have been proposed and implemented \cite{Axon18}.

The prioritisation of security requirements may vary across contexts and types of information. For example, time-critical safety contexts will prioritise having the information availability quickly enough for the UTM AI algorithms to make safe decisions, whereas in other cases competing stakeholders may prioritise the confidentiality of their commercially sensitive information. Therefore, a Blockchain-based system used in this context may need to be adaptable, not only in light of practical factors such as potential temporary wireless networking constraints (see Section II), but also based on security-requirement prioritisations, which may need to be back-propagated from the decision-making algorithms (this is reviewed next in more detail in Section IV).

\begin{figure*}[t]
     \centering
     \includegraphics[width=0.9\linewidth]{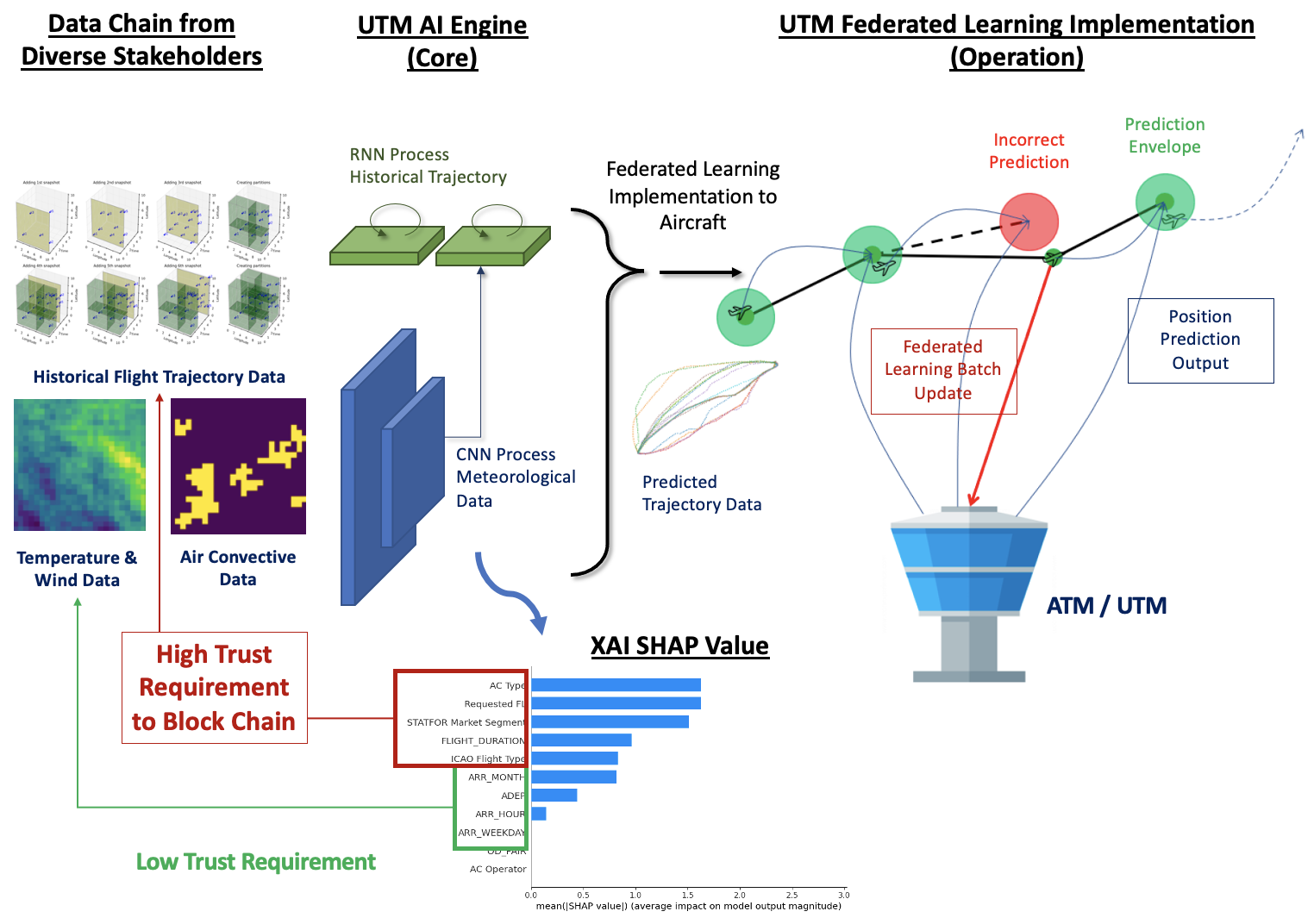}
     \caption{Case Study on Flight Trajectory and Arrival Time: arranged in format of Fig.\ref{fig1}. (top left) Diverse data pipelines from different stakeholders (airline, trajectory, meteorological) is used to predict aircraft position using a federated CNN \& RNN architecture. blockchain secures the data distribution with different consensus types based on the wireless channel. (bottom) Explainable AI (XAI) insight using various feature activation map and SHAP values can highlight the importance of certain data streams \cite{Dimitri21}. This informs new requirements for the blockchain to maintain or change the consensus distribution.}
     \label{fig6}
\end{figure*}

\section{XAI for Air Traffic Management with Backpropagation of Security Requirements}
\label{section4}

A wide range of tasks are performed by ATM to ensure the safe shared usage of air space. With increased autonomous drone traffic, often in non-segregated air space, there is active AI research in the following areas to enable UTM. There is already active AI research in these areas, some of which are safety critical, whilst others are also mission critical (e.g., vital for efficient and profitable aviation). A review of the UTM methods along with their corresponding blockchain consensus types (see Section II) and security requirements can be found in Table I. The gaps in existing literature is that most current usage of deep learning based air traffic management is not explainable \cite{EASA, GuoXAI20}. As such, we do not know which data inputs need greater scrutiny, security, or trust - vital for mission and safety critical mass autonomy \cite{LiTrust20}. By advancing in explainable AI (XAI), we can enable backpropagation of security and trust requirements to improve the accuracy and robustness of AI outputs for safety and critical air traffic management.

\subsection{Towards Explainable \& Trustworthy AI Metrics}

Globally implementation of AI in a variety of industries has raised the attention of legal issues regarding both its reliability (e.g., variability and robustness of performance to diverse circumstances) and the provenance of reasoning (e.g., which data features caused which decisions) \cite{EASA}. Machine learning techniques range from statistical classifiers to deep models (Recurrent Neural Network (RNN), and Convolutional Neural Network (CNN)). Indeed, these algorithms can be federated to local airspace or air platform operations, further highlighting the importance of distributed consensus adaptation. According to emerging national and international legal requirements (e.g., EU regulatory framework on artificial intelligence), there is an urgent requirement for developing methods and KPIs for both AI explainability and trust \cite{LiTrust20}. Explainability in AI (XAI) is extracted from different characteristics of machine learning models with multi-level expression of reasoning in statistics, semantics, symbolic expression, and logical reasoning. 

\subsubsection{Data Feature Based XAI:} Data is the most direct, meaningful, and detailed explanation in ML models, the raw explanation highlights the features with high contribution to the decision-making, which contains rich unbiased and unmodified raw information, but lacks expressions in logic (why and how extracted features cooperated). At the perhaps most intuitive level of explainability, one can post-hoc visualise the features that are important based on their weights or gradients of local nodes in the NN after training. In a gradient-based approach, we calculate the gradient of each input feature with respect to an output, where a small change in the input data feature leads to the level of outcome change can be visualised (e.g., DeepLIFT). Local features in hidden layers are non-linear and therefore the interpretation maybe not trivial. Here, approaches such as Shapley Additive Explanations (SHAP) can play a role to identify the average marginal contribution of a data features across all possible coalitions. 

\subsubsection{Summative XAI Models:} Summative explanation using design explainable models and surrogate models (high transparency `white box') mentioned above, generates smoothing and fitted explanations based on the processing of statistics from low transparency black box. For example, the decision flow of on-board autonomous model is not visible from Raw Explanation explanations, especially for CNN encoded high dimensional features, while summative explanation could generate a fitted symbolic expression of the DL model to clarify the relationships among inputs in formula expressions \cite{GuoXAI20}:
\begin{itemize}
    \item Global model level: (1) the Meijer G-function is a general function intended to include most known special functions and classes. As such, the Meijer G-function provides a flexible framework to discover the mapping between input variables and output solution. (2) Topological data analysis (TDA) presents an opportunity to examine how features are connected together via a graph, which can generate graph measures for feature cliques.
    \item Local model level: Local interpretable model-agnostic explanations (LIME) introduces a measure of complexity, such that one solves the following to obtain the minimum explanation, where we measure how unfaithful a reduced model is in the locality of a data point.
\end{itemize}
Examples of ATM algorithms they can be deployed for is presented in Table 1.

\subsubsection{Transparency \& Trust in AI Models that Inform blockchain Security Requirements:} Human trust of data and AI is premised often on 2 categories: physical trust (e.g., XAI performance with secure and verified data - see above), and emotional trust (e.g., familiarity) \cite{LiTrust20}. Here, we draw on the aforementioned XAI models to inform both the physical trust (data and its ability to influence our AI model) and emotional trust (measure a human operator's sentiment). For example, input flight data that the AI algorithm finds to be more sensitive to, less transparent, and humans trust less will need greater security requirements. Therefore, the miscellaneous trust and algorithm performance metrics need to be transformed into blockchain security requirements. This is then back propagated to the blockchain controller (see Section II-III) to maintain or change the consensus distribution - see Fig.\ref{fig6}(bottom).

\subsection{Case Study: Flight Trajectory and Delay Prediction with Federated Learning}
We are applying AI to predict the trajectory, as well as the arrival delay. Diverse data from airlines and the environment is used to inform the AI engine. We mainly build a RNN that ingests previous historical aircraft trajectory and general knowledge of the aircraft and airline. Example data including aircraft type, cruise altitude, carrier market segment, time of month, week day and arrival hour, and airline operator. We then inform this with additional meteorological data (temperature, wind, air convective patterns) which is processed by a parallel bank of CNNs, and this feeds into the aforementioned RNN structure. The data pipelines use the blockchain consensus distribution method detailed in Sections II-III, using a range of PoX and CFT for this mission critical application. The RNN \& CNN architecture is then federated into separate aircraft to make local predictions, whilst minimizing the data exchange with the ground-based ATM/UTM provider.

Whilst deep learning is inherently less explainable, we can use a variety of emerging XAI methods (see Fig.\ref{fig6}(bottom)) \cite{Dimitri21} to show key features driving predictions, providing better understanding of the model and also areas that need operational improvements. We see that the SHAP XAI analysis indicates sensitive to: (1) aircraft type, (2) requested flight cruise altitude level, and (3) market segment. This indicates that some data chains may need to improve their blockchain consensus type from PoX to CFT or BFT, depending on the wireless network available. These requirements will be backpropagated to the blockchain controllers explained in Sections II and III.

\section{Open Challenges}
\label{section5}

We are at the intersection of two major drivers for ATM transformation – the need for its modernisation to cater for the ever-increasing demand in air traffic, and the need for increased automation and autonomy to integrate the new UAS entrants, without adding to the already saturated workload of human operators. In this review, we examined the development of aviation mission-aware blockchains to distribute consensus. Decentralised machine learning is used to generate locally adaptable and ultra-reliable solutions for mission critical applications such as collision avoidance. When connected together with AI-enabled UTM, this new digital ecosystem blueprint is tailor designed for mission and safety critical autonomy sectors. Yet, it still requires significant research and development by the academic and industrial community to ensure it can enable future air and autonomous mobility. 

Future research challenges in this area include:
\begin{enumerate}
    \item High reliable and low latency consensus (at MAC and higher layers of network): critical for the ATM/UTM systems for real-time decisions.
    \item Networked privacy-preserving tracking: integrating blockchains that enable secure Self Sovereign Identity (SSI) for UAS threat identification. 
    \item Cognitive UTM: The transition towards Performance-Based Regulations (PBR) and Performance-Based Operations (PBOs) (see EASA PBE 2014, FAA PBO 2018 reports) aim to enable integration of more autonomy, including regulatory approval of AI solutions \cite{Dimitri21}.  
\end{enumerate}
We acknowledge that as we go forwards, the aeronautical data services will be more diverse, complex and dynamic than what is available today, and these will have to be accommodated whilst still achieving and even improving the level of integrity and reliability of existing conventional operations. Safety critical information will be needed at a much higher fidelity than in today’s solutions.

\section*{Acknowledgements}
We would like to thank the Fly2Plan project (70883) industrial (Heathrow Airport, National Air Traffic Services, SITA, International Airlines Group, Rockport, Altitude Angel, SCirium, IBS, Consortiq and the Digital Catapult) as well as research contributors (Mr. A. El Youmi) that provided data and conducted research far beyond the scope of this review.

\bibliographystyle{IEEEtran}
\bibliography{Ref}

\end{document}